\documentclass[hyper]{JHEP}
\pdfoutput=1
\usepackage[latin1]{inputenc}
\usepackage{graphicx,amssymb,bm,latexsym,amsmath}
\usepackage{epstopdf,subcaption}
\usepackage{amsfonts}
\usepackage{cite}
\usepackage{commath}

\title{$N$-spike string in $AdS_3 \times S^1$  with mixed flux}
\author{Rashmi R. Nayak, Priyadarshini Pandit, Kamal L. Panigrahi\\
	 Department of Physics, Indian Institute of Technology Kharagpur, Kharagpur 721 302, India\\
	Email: \email{rashmi.string@gmail.com,pandit006@iitkgp.ac.in, panigrahi@phy.iitkgp.ac.in}}

	\abstract{ Sigma model in $AdS_3\times S^3$ background supported by both NS-NS and R-R fluxes is one of the most distinguished integrable models. We study a class of classical string solutions for $N$-spike strings moving in $AdS_3 \times S^1$ with angular momentum $J$ in $S^1 \subset S^5$ in the presence of mixed flux. We observe that the addition of angular momentum $J$ or winding number $m$ results in the spikes getting rounded off and not end in cusp. The presence of flux shows no alteration to the rounding-off nature of the spikes. We also consider the large $N$-limit of $N$-spike string in $AdS_3 \times S^1$ in the presence of flux and show that the so-called Energy-Spin dispersion relation is analogous to the solution we get for the periodic-spike in $AdS_3-pp-$wave $\times S^1$ background with flux.}
	
	\begin{document}
	\section{Introduction and Summary}
	AdS/CFT correspondence \cite{Maldacena,Witten:1998qj,Gubser:1998bc} proposed by J. Maldacena, states that string theory on asymptotically $AdS$ background in $d$ dimensional spacetime is equivalent to Conformal Field Theory (CFT) living on its ($d-1$) dimensional boundary. It has been a striking discovery in the past two decades, unfolding various exciting possibilities to address the non-trivial strongly coupled dual gauge field theory in the planar limit as it seems duality is more tractable in that limit. Manifesting duality in general for all values of parameters has been highly challenging. However, restricting the domain of the correspondence to large charge limit has made the problem much simpler. The most studied example of AdS/CFT correspondence is the duality between $\mathcal{N}=4 $ Super Yang-Mills (SYM) theory and type IIB superstrings in $AdS_5\times S^5 $ background. However, in order to establish the duality in different sub-sectors of the theory on both sides of the correspondence, multiple attempts have been made in the past few years. The semi-classical approximation is considered one of the most appropriate methods used to find the spectrum in different target space geometries. Anomalous dimensions of the dual boundary field theory operators \cite{GKP,Berenstein:2002jq,Frolov:2003qc,Frolov:2003uk,Frolov:2003tu,Tseytlin:2003ii,Engquist:2003rn,spiky1} can be extracted from the dispersion relation among different conserved charges of certain rigid semi-classical string trajectories in AdS space using this approximation method. Various classical string solutions have been studied in order to understand the spectrum of $AdS_5\times S^5$ superstring theory in, e.g, \cite{Kruczenski:2008bs,Ishizeki:2008tx,Jevicki:2008mm,Jevicki:2009uz,Freyhult:2009bx,Kruczenski:2010xs,Losi:2010hr,Banerjee:2015nha,Banerjee:2019puc}. \\
	The other most intensively studied example of the gauge-gravity duality in lower dimension is $AdS_3/CFT_2$ \cite{Sfondrini,VIII,David:2010yg,SST,AB,SW} duality. It claims that type IIB string theory on $AdS_3\times S^3\times M^4$ is equivalent to the ${\cal N} = (4,4)$ superconformal field theory in three dimensions with $M^4$ being $T^4$ \cite{Ads3S3T4A,Borsato:2013hoa,BianchiHoare,BothM4,completeworldsheet,Ads3S3T4C,BMN mismatch,Ads3S3T4D,BPS} or $S^3\times S^1$ \cite{Rughoonauth,Borsato:2015mma,Abbott,Abbott:2013mpa,EGGL,XIV,Tong,EberhardtGopakumar,Pittelli:2017spf,BSZ,wulff,review,Moo}. In recent years,  the $AdS_3 \times S^3$ background supported by both R-R and NS-NS fluxes ($H_3 = dB_2$ and $F_3 = dC_2$) has been shown to be integrable \cite{Cagnazzo:2012se,Wulff:2014kja,HT,B.Hoare,Lloyd}. There have been extensive studies on $S$-matrix \cite{Modulispace} along with the integrable structure of the corresponding theory. Aiming to comprehend the duality better,  a large number of classical string solutions have been constructed in this background. The background solution has been shown to satisfy type IIB supergravity field equations, given the parameters are related by the constraint $q^2+ {\hat q}^2= 1$, where $q$ and  $\hat{q}$ are associated to the field strength of NS-NS fluxes and RR fluxes respectively. There have been multiple studies on various classical string solutions in detail in \cite{FinitegapBabichenko,masslessfinitegap,HN1,Rotating1,C.Ahn,both,HN2,Hernandez:2018gcd,Biswas:2020wfn}.  We study class of rigidly rotating strings and speculate its corresponding dual states in gauge theory in the mixed flux background. The Energy-Spin dispersion relation of the folded spinning string or GKP string solution in $AdS_3$ was analyzed in \cite{GKP}. This relates to the anomalous dimension of the single trace, twist-two dual gauge theory operators in the $SL(2)$ sector of the SYM theory in the large spin ($S$) limit given by $\mathcal{O}=$Tr$(\Phi \nabla^S_+ \Phi)$. The above closed rigid folded spinning string solution has been generalized to the case of $N$-spike strings in $AdS_3$ \cite{spiky1} whose dual gauge theory operator has been identified by certain higher twist operator of the form Tr$(\nabla^{S_1}_+ \Phi_1\nabla^{S_2}_+ \Phi_2...\nabla^{S_n}_+ \Phi_n)$. The spikes or cusps, which can be considered as the derivative discontinuity on a 2-dimensional worldsheet, approach the boundary of $AdS_3$ in the limit $S\to \infty$. A detailed study was done in \cite{Kruczenski:2008bs}, and found that the solution of spikes in this large $S$ limit corresponds to a string solution in an $AdS_5-pp-$ wave metric 
	\begin{equation}
	ds^2=\frac{1}{z^2}\left[2dx_+dx_- -\mu^2 (z^2+x_i^2)dx_+^2+dx_idx_i+dz^2\right],~~~~~~~i=1,2
	\label{0}
	\end{equation}
	It can be seen here that, we get the above metric (\ref{0}) by zooming at the near boundary region of $AdS_5$ while simultaneously moving close to the speed of light in the angular direction. In turn there is a correspondence between string theory in an $AdS_5-pp-$ wave $\times $ $S^5$ and $\mathcal{N}=$ 4 SYM theory in a 4-dimensional $pp-$wave background \cite{Berenstein:2002jq}. The dual metric takes the following form
	\begin{equation}
	ds^2_{ft}= 2dx_+dx_--\mu^2x_i^2 dx_+^2 + dx_idx_i
	\label{01}
	\end{equation} 
	This conformally flat metric is the boundary metric of the spacetime defined in  eq (\ref{0}). Numerous string solutions moving in the $AdS_3-pp-$wave background were studied, e.g., in \cite{Ishizeki:2008dw}. Here in this article, the solutions we find may be interpreted as the spiky string solutions in the presence of mixed flux, which corresponds to spikes extending to the boundary while the number of spikes approaching infinity $N\to \infty$ keeping $\frac{E+S}{N^2}$ and $\bar{\gamma}=\frac{E-S}{N}$ fixed. All spikes (here infinite in number) contributes a finite amount $\bar{\gamma}$ to the anomalous dimension $\gamma=N \bar{\gamma}=E-S$ exactly similar to the case studied in \cite{Ishizeki:2008tx}. It can be argued that this large $N$ limit of the spiky string was a meaningful thermodynamic limit of a certain spin chain, which defines strong-coupling asymptotics of the corresponding anomalous dimension.
	The Energy-Spin scaling relation may be written as  \\
	\begin{equation}
	E=S+J+\gamma(S,J,m,N;\lambda)
	\end{equation}
	where $\lambda$ is t' Hooft coupling. 
	We study the $N$-spike string solution in mixed flux background discussed in \cite{Banerjee:2019puc} in the presence of the angular momentum $J$ (and winding m) in a maximal circle $S^1$ in $S^5$. We use conformal gauge and embedding coordinates as discussed in \cite{Jevicki:2008mm,Kruczenski:2006pk,Arutyunov:2003uj,Frolov:2003qc} for finding the rigid rotating spiky string solution. The solutions we get are almost similar to those in \cite{Hayashi:2007bq}. We observe that the shape of the string in the $AdS_3$ space with mixed flux is identical to the original spiky string with $\mathcal{J}=m=0$. However, a more detailed analysis of the string profile shows that for non-zero $\mathcal{J}$, the spikes get rounded-off instead of ending in cusp. Similar behaviour was noticed in \cite{Ishizeki:2008tx}. We also discuss $pp-$wave limit to the $N$-spike string solution in $AdS_3\times S^1$ with NS-NS flux. This limit corresponds to taking $N\to \infty $ for non-zero $\mathcal{J}$. We find that the polynomial we get here by virtue of this limit show complete resemblance to the expression we get for the periodic spike in $AdS-pp-$wave $\times S^1$.\\
	
	The rest of the paper is organized as follows. In section 2, we consider the large $N$ limit of the spiky string moving in $AdS_3$ background with mixed flux and expression for conserved charges are found. It was shown later that exactly the same result could be obtained by considering a string moving in an $AdS-pp-$wave background with flux. The corresponding spacetime can be achieved in particular by taking a limit of $AdS_5$, i.e., zooming at its near boundary region and moving simultaneously close to the speed of light in the angular direction ($S \to \infty$). Section 3 incorporates two subsections. In the first part, we generalize the study of the rigid $N$-spike string solution in $AdS_3$ background with mixed flux to the addition of non-zero angular momentum $J$ and winding number $m$ in $S^5$  and show that the spikes do not end in cusps instead get rounded off. This extrapolates the result obtained in \cite{Ishizeki:2008tx}. For smaller values of $q$, the profile is being depicted. The second subsection deals with two separate cases, i.e., $J=0$ in the presence of mixed flux and the $pp-$wave limit to $N$-spike string in $AdS_3 \times S^1$ with flux. In section 4, we generalize the result obtained in section 2. We study  periodic spike solution in $AdS-pp-$wave $\times S^1$ background in the presence of NS-NS flux and find its connection to the $N$-spike string in $AdS_3 \times S^1$ with flux in large $N$ limit. In section 5, we present our conclusions and outlook. \\

	\section{Large `$N$' limit of $N$-spike string in mixed flux background}
	In this section we consider large `$N$' limit (N$\to \infty$) of the $N$-spike string solution in $AdS_3$ with mixed flux \cite{Jevicki:2008mm}.
	
	\subsubsection*{  A. Spiky string solution with large $N$ limit}
	To start with, we consider rigid $N$-spike string \cite{spiky1} rotating in $AdS_3$ background in presence of NS-NS and R-R fluxes. The expression for the background metric along with fluxes are as follows
	\begin{equation}
	ds^2_{AdS_3} = -\cosh^2\rho dt^2+d\rho^2+\sinh^2\rho d\theta^2,~~
	~~~
	b_{t\theta}=q\sinh^2\rho,~~~c_{t\theta} = \sqrt{1-q^2}\sinh^2\rho.
	\end{equation}
	We consider the following ansatz
	\begin{equation}
	t=\tau,~~~~~\theta=\omega \tau +\sigma,~~~~\rho=\rho(\sigma)
	\end{equation}
	where $q$ is the flux parameter whose value ranges from $0\leq q\leq1$.  To describe the motion of a rotating spiky string in this background, we follow the discussion in \cite{Kruczenski:2006pk}. In order to find the solution of such string rotating in  $AdS_3\times S^3$ background with mixed flux, we consider Nambu-Goto action given by 
	\begin{equation}
	S_{NG}=\frac{\sqrt{\lambda}}{2\pi}\int d\tau d\sigma\Big[\sqrt{-(\dot{X}^2 X^{\prime 2}-(\dot{X} X^\prime)^2)}-\frac{\epsilon^{ab}}{2}B_{mn}\partial_a X^m\partial_bX^n\Big],
	\end{equation}
	where dot indicate derivative with respect to $\tau$ and prime with respect to $\sigma$. $\lambda$ is the t'hooft coupling constant and $\epsilon^{ab}$ is the antisymmetric tensor with $\epsilon^{01}=1=-\epsilon^{10}$. We can see that $\rho$ satisfies 
	\begin{equation}
	\frac{dv}{d\sigma}=(1-v^2)\sqrt{\frac{v(1-v_1)}{v-v_1}}\frac{\sqrt{(v_0^2-v^2)-q^2v_0^2(1-v)^2-2qvv_0(1-v)\sqrt{1-v_0^2}}}{(v\sqrt{1-v_0^2}+qv_0(1-v))},
	\end{equation}
	where $v_0$ and $v_1$ corresponds to the positions of the spikes and valleys, respectively, which can also be interpreted as the maximum and minimum value of $\rho$ or $v$. Now we will construct Noether charges by considering the isometries of the given background. The expression for the angular separation between the spikes, Spin, and Energy are as follows
	\begin{eqnarray}
	\Delta\theta=\frac{2\pi}{N}=2 \int_{v_1}^{v_{max}} \sqrt{\frac{v-v_1}{v(1-v_1)}}&& \frac{v\sqrt{1-v_0^2}+qv_0(1-v)}{\sqrt{(v_0^2-v^2)-q^2v_0^2(1-v)^2-2qvv_0(1-v)\sqrt{1-v_0^2}}}\nonumber\\&&\times \frac{dv}{1-v^2},
	\end{eqnarray}
	
	\begin{equation}
	\frac{\mathcal{S}}{N}=\frac{\sqrt{1+v_1}}{4\pi v_0}\int_{v_1}^{v_{max}}\frac{dv}{v(1+v)}\frac{\sqrt{{v_0^2-v^2-q^2v_0^2(1-v)^2-2qv_0v(1-v)\sqrt{1-v_0^2}}}}{v(v-v_1)},
	\end{equation}	
	
	\begin{eqnarray}
	&&\frac{\mathcal{E}}{N}= \frac{\sqrt{1-v_1}}{4\pi v_0}\int_{v_1}^{v_{max}}\frac{dv}{v(1-v)}\frac{\sqrt{v_0^2-v^2-q^2v_0^2(1-v)^2-2qv_0v(1-v)\sqrt{1-v_0^2}}}{\sqrt{v(v-v_1)}}\nonumber\\
	&+&\frac{1}{2\pi v_0\sqrt{1-v_1}}\int_{v_1}^{v_{max}}\frac{dv \sqrt{v(v-v_1)}}{v^2(1-v^2)}\frac{(v\sqrt{1-v_0^2}+qv_0(1-v))^2}{\sqrt{v_0^2-v^2-q^2v_0^2(1-v)^2-2qv_0v(1-v)\sqrt{1-v_0^2}}}\nonumber\\
	&-&\frac{q}{2\pi \sqrt{1-v_1}}\int_{v_1}^{v_{max}}\frac{dv}{v^2(1+v)}\frac{\sqrt{v(v-v_1)}\left(v\sqrt{1-v_0^2}+qv_0(1-v)\right)}{\sqrt{v_0^2-v^2-q^2v_0^2(1-v)^2-2qv_0v(1-v)\sqrt{1-v_0^2}}},\nonumber\\
	\end{eqnarray}
	where $N$ represents the number of spikes, and the expression for Energy and Spin is written as $E=2\pi T\mathcal{E} $ and $S=2\pi T\mathcal{S}$  respectively, where $T=\frac{\sqrt{\lambda}}{2\pi}$ is the tension of the string. We could have written the resulting expression for the above integrals in terms of the standard elliptic functions but for taking the limit we are interested in, it is convenient to keep these integral representations intact.\\
	Let us consider the following scaling
	\begin{equation}
	v_0\to\epsilon^2v_0,~~~~~~v_1\to\epsilon^2v_1,~~~~~~\epsilon\to 0
	\end{equation}
	It can be seen that under such scaling, the following expressions scale as
	\begin{equation}
	N\sim\frac{1}{\epsilon^2},~~~~~\frac{\mathcal{E+S}}{N^2}\sim\frac{1}{\epsilon^2},~~~~~\frac{\mathcal{E-S}}{N}\sim finite.
	\end{equation} 
	Now we compute the following finite quantities $\cfrac{\mathcal{E-S}}{N}$ and $ \cfrac{\mathcal{E+S}}{N^2}$ under the same limit which we took above. The expression has been given in Appendix 1 in eqn(\ref{20}),(\ref{21}). It can be seen from the expression that 
	 they are invariant under the above scaling and depend only on the ratio $$b\equiv \frac{v_1}{v_0}.$$ 
	This dependence can be seen clearly from the following expression,
	\begin{eqnarray}
	\bar{\gamma}&\equiv&\frac{E-S}{N}=\frac{T}{\sqrt{(1-q)(1+b+q)}}\Bigg[(1-q)(1+b+q)~\mathbf{E}(p)\nonumber\\
	&+&(b+2q)~b~\mathbf{\Pi}\left(\frac{1-q-b}{1-q},p\right)-(1-q)\left(b+2(1+q)\right)\mathbf{K}(p)\Bigg],
	\label{14}\end{eqnarray}	
	\begin{eqnarray}
	\bar{P_-}\equiv\frac{E+S}{N^2}&=&\frac{2T}{b \pi (1-q)(1+q+b)}\Bigg[(1-q)(1+b+q)\mathbf{E}(p)\nonumber\\&-&b(1+q)\mathbf{K}(p)- b^2\mathbf{\Pi}\left(\frac{1-q-b}{1-q},p\right)\Bigg],
	\label{15}\end{eqnarray}
	where $$p=\sqrt{\frac{(1+q)(1-q-b)}{(1-q)(1+q+b)}}.$$
	K(p), E(p) and $\Pi\left(\frac{1-q-b}{1-q},p\right)$ are the standard elliptic functions of first, second and third kind respectively. It can be clearly seen here that the expression for $\bar{\gamma}$ can be written as a function of $\bar{P_-}$. 
	From gauge/gravity duality, the correspondence between the energy spectrum of semiclassical string states in the large charge limit and the anomalous dimension of single trace operators from $SL(2)$ sector of dual boundary field theory was established. One of the most interesting example in support of this argument is given by folded spinning strings in \cite{GKP}, which deals with twist-two operators in gauge theory side and folded spinning string in string theory side in AdS space. A generalization to this has been shown in \cite{spiky1}, where a higher twist operator is found to be dual to string solutions with N-spikes. Furthermore, there is a remarkable observation that such operators can be described by the Heisenberg spin chain with $N$ denoting the number of sites \cite{Belitsky:2006en}. The following correspondence suggests that the large $N$ limit under consideration should be a meaningful thermodynamic limit of the spin chain. We can reproduce the similar expression for $\bar\gamma$ and $\bar{P}_-$, when we proceed from the spin chain side describing the anomalous dimension of the dual field theory in the strong coupling limit.
	
	\subsubsection*{  B.  Spiky string in $AdS-pp-$wave background with flux}
	Here we show that similar expressions as in (\ref{14}) and (\ref{15}) can be retrieved by taking the solution of the string moving in an $AdS-pp-$wave background into account. It can be claimed here that this metric contains the information in order to comprehend the thermodynamic limit of spin chain taken into consideration. This background which is found to be a particular limit of $AdS_5$ space, is a result of zooming in the near boundary region while simultaneously moving close to the speed of light along the angular direction. 
	\begin{enumerate}
		\item[1.] \it{String moving in AdS-pp-wave background with flux}
	\end{enumerate}
	We will reproduce $AdS-pp-$wave metric from the global $AdS_5$ spacetime. The global $AdS_5$ metric and the NS-NS flux can be written as 
	\begin{equation}
	ds^2=-\cosh^2\rho dt^2+ d\rho^2 + \sinh^2\rho \frac{(1+\frac{x_i^2}{4})^2 d\theta^2+dx_i dx_i}{(1+\frac{x_i^2}{4})^2},
	\end{equation}
	\begin{equation}
	B_{t\theta}= q \sinh^2\rho ~~dt\wedge d\theta,
	\end{equation}
	where $x_1,x_2$ and $\theta$ parametrize the 3-sphere. We introduce new set of coordinates $(z,x_+,x_-)$ which are related to the global coordinates as 
	\begin{equation}
	\rho=-\ln(2z),~~~~t=x_+-x_-,~~~~\theta=x_++x_-,
	\end{equation}
	now we rescale the newly introduced coordinates by a parameter $\epsilon$ as
	\begin{equation}
	x_+\to \mu^{-1}x_+,~~~~x_-\to 8\mu\epsilon^2 x_-,~~~~x_i\to 4\epsilon x_i,~~~~z\to\epsilon z.
	\end{equation}
	Taking the limit $\epsilon \to 0$ and keeping only the leading $\epsilon$ independent terms we found the metric as well as the two form B-field to be
	\begin{equation}
	ds^2=\frac{1}{z^2}\left[2dx_+dx_--\mu^2(z^2+x_i^2)dx_+^2+dx_idx_i+dz^2\right],
	\end{equation}
	\begin{equation}
	B_{x_+x_-}= \frac{q}{z^2}~ dx_+\wedge dx_-.
	\end{equation}
	We presume $x_i=0$ and $x_+=\tau$. Now the string action can be expressed as, 
	\begin{equation*}
	S_{NG}=-\frac{\sqrt{\lambda}}{2\pi}\int d\tau d\sigma\Big[\sqrt{-(\dot{X}^2 X^{\prime 2}-(\dot{X}.X^\prime)^2)}-\frac{\epsilon^{ab}}{2}B_{mn}\partial_a X^m\partial_bX^n\Big].
	\end{equation*}	
	\begin{equation}
	S=-\frac{\sqrt{\lambda}}{2\pi}\int d\tau d\sigma \frac{1}{z^2}\sqrt{x_{-}^{\prime 2}\dot{x}_+^2 +2x^\prime_-\dot{x}_+\dot{z}z^\prime-2\dot{x}_-\dot{x}_+z^{\prime2}+\mu^2z^2z^{\prime 2}\dot{x}_+^2}-\frac{q}{z^2}x_-^\prime\dot{x}_+,
	\end{equation}
	The equation of motion for $x_-$  and $z$ can be expressed as 
	\begin{equation}
	-\partial_\tau \left(\frac{z^{\prime 2}}{z^2 M}\right)+\partial_\sigma\left(\frac{x_-^\prime+\dot{z}z^\prime}{z^2 M}-\frac{q}{z^2}\right)=0
	\end{equation}
	\begin{equation}
	\partial_\sigma\left(\frac{x_-^\prime\dot{z}-2\dot{x}_-z^\prime+\mu^2z^2z^\prime}{z^2M}\right) + \partial_\tau\left(\frac{z^\prime x_-^\prime}{z^2M}\right)=-\frac{2}{z^3}(M-q)+\frac{\mu^2z^{\prime 2}}{zM},
	\end{equation}
	where$$M=\sqrt{{x_{-}^{\prime 2} +2x^\prime_-\dot{z}z^\prime-2\dot{x}_-z^{\prime2}+\mu^2z^2z^{\prime 2}}}$$
	The conserved momenta takes the form
	\begin{equation}
	P_+=\int d\sigma\frac{\partial \mathcal{L}}{\partial\dot{x}_+}=-T\int\frac{d\sigma}{z^2 M}(\mu^2z^{\prime2} z^2 +\dot{z}z^\prime x_-^\prime-\dot{x}_-z^{\prime 2} +x_-^{\prime 2})-\frac{q}{z^2}x_-^\prime,
	\end{equation}
	\begin{equation}
	P_-=\int d\sigma \frac{\partial\mathcal{L}}{\partial\dot{x}_-}=T\int d\sigma\frac{z^{\prime 2}}{z^2M}.
	\end{equation}
	\begin{enumerate}
		\item[2.] \it{Periodic spike solution in $AdS-pp-$wave background}
	\end{enumerate}	
	We now consider  $x_-=\sigma$ in addition to the earlier gauge choice $x_+=\tau$ and obtain the solution for string moving with constant velocity along $x=\frac{1}{\sqrt{2}}(x_++x_-)$.
	\begin{equation}
	z=z(\xi),~~~~~\xi\equiv x-vt=x_+-\frac{v+1}{v-1}x_-=\tau-\frac{1}{\eta_0^2}\sigma, ~~~~~ \text{with}~~\eta_0^2\equiv\frac{v-1}{v+1} .
	\end{equation}
	Here we consider the solution for which $v>1$.  This claims that the string remains in the bulk and does not reach the boundary. Taking the above ansatz into consideration, we find the equation for motion of $x_-$ to be
	\begin{equation}
	\partial_\xi\left(\frac{1+qM}{\eta_0^2z^2M}\right)=0
	\end{equation} 
	\begin{equation}
	\implies \partial_\xi z=\frac{\eta_0^2}{\mu (z^2+qz_0^2)}\sqrt{\frac{(z^2+(1+q)z_0^2)((1-q)z_0^2-z^2)}{z^2-z_1^2}}
	\end{equation}
	where $z_0$ is a constant and $\mu z_1\equiv \sqrt{2} \eta_0$.
	The conserved quantities can be computed in terms of standard elliptic integrals as (the intermediate step is given in eqn \ref{22} and \ref{23}) of Appendix 1.

	\begin{eqnarray}
	P_-=\frac{2T}{\mu z_0^2}\frac{1}{b\sqrt{(1-q)(1+b+q)}}&&\Bigg[(1-q)\left(1+b+q\right)\mathbf{E}(p)-(1+q)~b~\mathbf{K}(p)\nonumber\\&-&b^2~\mathbf{\Pi}\left(\frac{1-q-b}{1-q},p\right)\Bigg],
	\label{16}\end{eqnarray}
	\begin{eqnarray}
	P_+= \frac{T\mu}{\sqrt{(1-q)(1+q+b)}}&&\Bigg[(1-q)\left(1+q+b\right)\mathbf{E}(p)+b~(2q+b)~\mathbf{\Pi}\left(\frac{1-q-b}{1-q},p\right)\nonumber\\&-&(1-q)\left(b+2(1+q)\right)\mathbf{K}(p)\Bigg],
	\label{17}\end{eqnarray}
	where$$p=\sqrt{\frac{(1+q)(1-q-b)}{(1-q)(1+q+b)}},~~~~~~~~b=\frac{z_1^2}{z_0^2}$$.
	Putting $b \to 0$ we get,
	\begin{equation}
	P_-\simeq \frac{2T}{\mu z_0^2}\frac{\sqrt{1-q^2}}{b},~~~~~~~~~P_+~\simeq \mu T \ln \frac{b}{1-q^2}
	\end{equation}
	so that $$P_+\simeq -\sqrt{1-q^2}\mu T \ln(P_-).$$
	The multiplicative factor $\sqrt{1-q^2}$ along with the logarithmic behaviour at the leading order in the large spin limit i.e.,  $S\to\infty $ was anticipated from the result we obtain for the Energy-Spin dispersion relation in \cite{Banerjee:2019puc}. The expression for the angular separation $\Delta x_-$ (for constant $x_+$) between spikes is found to be
	\begin{eqnarray}
	\Delta x_-=\frac{\mu z_0^2}{\sqrt{(1-q)(1+b+q)}}\Bigg[(1-q)\left(1+b+q\right)\mathbf{E}(p)-b(1+q)\mathbf{K}(p)-b^2\mathbf{\Pi}\left(n,(p)\right)\Bigg].\nonumber\\
	\end{eqnarray}
	This gives 
	\begin{equation}
	P_-\Delta x_- = \frac{2T}{b(1-q)(1+b+q)}\Bigg[(1-q)(1+b+q)\mathbf{E}(p)-(1+q)~b~\mathbf{K}(p)-b^2~\mathbf{\Pi}\left(n,p\right)\Bigg]^2
	\end{equation}
	\begin{eqnarray}
	P_+= \frac{\mu T}{\sqrt{(1-q)(1+q+b)}}\Bigg[(1-q)(1+q+b)\mathbf{E}(p)&-&(1-q)(b+2(1+q))\mathbf{K}(p)\nonumber\\&& + b~(2q+b)\mathbf{\Pi}\left(n,p\right)\Bigg].
	\end{eqnarray}
	where $$n= \frac{1-q-b}{1-q}.$$
	Now it can be argued that with proper substitution as in \cite{Ishizeki:2008tx} i.e, $P_-\Delta x_-=\pi \bar{P}_-,P_+=\mu \bar{\gamma}$, they exactly resembles the expressions we get in eq.(\ref{16}) and eq.(\ref{17}). So with this $AdS-pp-$wave metric, we can also determine about certain thermodynamics limits in spin chain.
	In this large $N$ limit i.e, $N\to \infty$ with $(E+S)/N^2$ being fixed, $(E+S)/N^2$ can be written as $\cfrac{E+S}{N}\times \Delta \theta$ where the number of spikes and the angle difference between the spikes are related to each other and scales as $\Delta \theta\sim \frac{1}{N}$. So, $\cfrac{E+S}{N^2}=\cfrac{E+S}{N}\times \Delta \theta$ converts to $P_-\Delta x_-$ in the $pp$- wave picture. Here $P_+$ and $P_-$ can be interpreted as $E-S$ and $-(E+S)$ respectively.\\
	
	\section{$N$-spike strings in $AdS_3 \times S^1$ with mixed flux}
	
	\subsection{General Solution}
	Here, in this section, we shall generalize the $N$-spike string solution found in $AdS_3$ in the presence of NS-NS flux to the case of addition of angular momentum $J$ and winding $m$ in $S^1$ in $AdS_3\times S^1$ background. The string sigma model for rotating closed semi-classical string in $AdS_3 \times S^3$ in the absence of flux reduces to Neumann-Rosochatius system, which is a 1-dimensional integrable model. This system describes a 3-dimensional harmonic oscillator on a sphere with centrifugal potential. The presence of flux introduces an additional term in the lagrangian of Neumann-Rosochatius system \cite{Arutyunov:2003uj}.
	
	We will start with the $AdS_3 \times S^1$ metric with mixed fluxes in embedding coordinates \cite{Kruczenski:2006pk}. The world-sheet action of the bosonic part in the conformal gauge is given by 
	\begin{equation}
	S = -\frac{\sqrt{\lambda}}{4\pi}\int d\tau d\sigma \left[ G_{mn}^{(AdS_5)}(y) \partial_a y^m \partial^a y^n + G^{(S^1)} \partial_a x \partial^a x -\epsilon^{a b} \partial_a y^M \partial_b y^N b_{M N}\right].  \label{1}
	\end{equation} 
	The metric of $AdS_3$ and $S^1$ and the two form B-field have the following standard form in terms of global coordinate;
	\begin{equation*}
	(ds^2)_{(AdS_3)} = -\cosh^2\rho dt^2 +d\rho^2 + \sinh^2\rho d\theta^2, \end{equation*}
	\begin{equation}
	(ds^2)_{S^1} = d\varphi^2
	\end{equation}
	\begin{equation}
	b_{t\theta}= q \sinh^2\rho ~~dt\wedge d\theta
	\end{equation} 
	It is convenient to represent (\ref{1}) as an action for the sigma model (we follow Neumann-Rosochatius (NR) 1-d integrable system).
	\begin{equation}
	S = \frac{\sqrt{\lambda}}{2\pi}\int d\tau d\sigma \left(L_{AdS} +L_S + \frac{\epsilon^{ab}}{2}\partial_a Y^M \partial_b Y^N B_{MN}\right)
	\label{2}
	\end{equation} 
	where
	\begin{equation}
	L_{AdS} = -\frac{1}{2}\eta^{MN}\partial_a Y_M \partial^a Y_N^* - \frac{1}{2}\Lambda(\eta^{MN} Y_M Y_N^* +1)
	\end{equation} 
	\begin{equation}
	L_S = -\frac{1}{2}\partial_a X \partial^a X^* - \frac{1}{2}\tilde{\Lambda}(X X^* -1),
	\end{equation}
	Here $Y_M$, $M=0,1,2,3$ and $X$ are the embedding coordinates of $R^4$ and $R^2$ with $\eta^{MN} =(-1,+1,+1,-1)$ in $L_{AdS}$ and with Euclidean metric in $L_s$ respectively. $\Lambda$ and $\tilde{\Lambda}$ are the Lagrange multipliers. We use conformal gauge constraints in order to supplement the action given in (\ref{2}). Now since we want to work in the Neumann-Rosochatius system, we need the relation between global coordinates and embedding coordinates which can be given as follows:
	\begin{equation}
	Y_0 = Y_0+iY_3 = \cosh\rho e^{it}, ~~Y_1 = Y_1+iY_2 = \sinh\rho e^{i\theta}, ~~X= X_1+iX_2 = e^{i\varphi}.
	\end{equation}
	In embedding coordinates, the string Lagrangian takes the form
	\begin{eqnarray}
	L= -\frac{1}{2}\Big[-\partial_a Y_0 \partial^a Y_0^* &+& \partial_a Y_1 \partial^a Y_1^* + \partial_a X \partial^a X^* + \Lambda(~\abs{Y_0}^2 - \abs{Y_1}^2 - 1) \nonumber \\&& + \tilde {\Lambda} (~\abs{X}^2 -1) -\epsilon^{ab}\partial_a Y^M \partial_b Y^N B_{MN} \Big]
	\end{eqnarray}
	The conformal constraints are
	
	\begin{equation}
	-\abs{\dot{Y_0}}^2 +\abs{\dot{Y_1}}^2 + \abs{\dot{X}}^2 - \abs{Y_0^\prime}^2 +\abs{Y_1^\prime}^2 +\abs{X^\prime}^2 = 0, ~~   -\dot{Y_0} Y_0^{\prime *} + \dot{Y_1} Y_1^{\prime *} + \dot{X} X^{\prime *} +c.c = 0
	\end{equation}
	We shall consider the following rigid string ansatz,
	
	\begin{equation}
	Y_0 = y_0(\xi)e^{i\omega_0 \tau}, ~~~Y_1 = y_1(\xi)e^{i\omega_1 \tau},~~~X = x(\xi)e^{i\omega_2 \tau}
	\end{equation} 
	where $$\xi=\alpha\sigma +\beta \tau.$$ We have taken generalised ansatz where $\xi$ is both $\sigma$ and $\tau$ dependent. 
	With this ansatz, the string lagrangian reduces to the one-dimensional integrable mechanical system  \cite{Arutyunov:2003uj,Kruczenski:2006pk}.
	
	The resulting expression on combining the two given conformal constraints can be written as
	\begin{equation}
	(\beta^2-\alpha^2)\big(\abs{y_0^\prime}^2 - \abs{y_1^\prime}^2 - \abs{x^\prime}^2\big) - \omega_0^2\abs{y_0}^2 +\omega_1^2\abs{y_1}^2+ \omega_2^2\abs{x}^2   = 0
	\end{equation}
	\begin{equation}
	\frac{\beta^2-\alpha^2}{2\beta}(-\omega_0 \zeta_0+\omega_1 \zeta_1 +\omega_2 \zeta_2)-\omega_0^2\abs{y_0}^2+\omega_1^2\abs{y_0}^2+\omega_2^2\abs{x}^2  = 0
	\end{equation}
	where
	\begin{equation}
	\zeta_0= i(y_0y_0^{*\prime}- y_0^\prime y_0^*),~~ \zeta_1= i(y_1y_1^{*\prime}- y_1^\prime y_1^*),~~ \zeta_2= i(x x^{*\prime}- x^\prime x^*)
	\end{equation}
	Here $y_0, y_1$ and $x$ are taken in general complex also $\abs{y_0}^2-\abs{y_1}^2=1$ and $\abs{x}^2=1$, so they may be expressed as
	$$y_0= r_0(\xi)e^{i\phi_0(\xi)},~~ y_1 = r_1 (\xi) e^{i\phi_0(\xi)}, ~~x=e^{i\psi(\xi)},~~ r_0^2-r_1^2=1$$
	So, we can write $$Y_0=r_0(\xi)e^{if},~~Y_1=r_1(\xi)e^{ig},~~ X=e^{ih}$$
	where$$ f=\phi_0(\xi)+\omega_0 \tau,~~g=\phi_1(\xi)+\omega_1 \tau,~~h=\psi(\xi)+\omega_2 \tau,$$
	With all the above substitutions, Lagrangian takes the following form 
	\begin{eqnarray}
	L&=& -\frac{1}{2} \Big[(\beta^2-\alpha^2)~(r_0^{\prime 2}- r_1^{\prime 2})+~ r_0^2~(\beta^2-\alpha^2)\left(\phi_0^{\prime}+\frac{\beta\omega_0}{\beta^2-\alpha^2}\right)^2-\frac{\alpha^2 r_0^2\omega_0^2}{\beta^2-\alpha^2}\nonumber\\&& - r_1^2(\beta^2-\alpha^2)\left(\phi_1^{\prime}+\frac{\beta\omega_1}{\beta^2-\alpha^2}\right)^2+\frac{\alpha^2r_1^2\omega_1^2}{\beta^2-\alpha^2}-(\beta^2-\alpha^2)\left(\psi^\prime+\frac{\beta \omega_2}{\beta^2-\alpha^2}\right)^2\nonumber\\&& +\frac{\alpha^2\omega_2^2}{\beta^2-\alpha^2}-2qr_1^2\alpha(\phi_1^\prime\omega_0-\phi_0^\prime\omega_1)+ \Lambda(r_0^2-r_1^2-1)\Big]
	\end{eqnarray}
	In order to find relevant closed string solution we need to impose the periodicity conditions
	\begin{equation*}
	r_0(\xi)=r_0(\xi + 2\pi\alpha),~~~~~ r_1(\xi)= r_1(\xi+ 2\pi\alpha),
	\end{equation*}
	\begin{equation*}
	\phi_0(\xi)= \phi_0(\xi + 2\pi\alpha) - 2\pi m_0,~ \phi_1(\xi)= \phi_1(\sigma + 2\pi\alpha) - 2\pi m_1,~ \psi(\xi)= \psi(\xi + 2\pi\alpha) - 2\pi m,
	\end{equation*}
	where $m_0,m_1,m$ are integer winding numbers. 
	The  Euler-Lagrange equations of motion for $\phi_0, \phi_1$ and $\psi$ can be calculated as
	\begin{eqnarray}
	\phi_0^\prime&=& -\frac{1}{\beta^2-\alpha^2}\left(\frac{A_0+q\alpha\omega_1 r_1^2}{r_0^2}+\beta\omega_0\right),~~~ \phi_1^\prime = \frac{1}{\beta^2-\alpha^2}\left(\frac{A_1}{r_1^2}-(q\alpha\omega_0+\beta\omega_1)\right),\nonumber\\&&\hspace{5.5cm} \psi^\prime= \frac{A_2-\beta \omega_2}{\beta^2-\alpha^2}
	\end{eqnarray}
	where $A_0, A_1, A_2$ are constants. In order to find out the winding number, we will integrate $\psi$ again to have a rotating string wound on a circle.
	\begin{equation}
	X=e^{i\varphi}, ~~~~~~~~~~ \varphi= \omega_2 \tau + \psi = \omega_2\tau -\frac{A_2-\beta\omega_2}{\beta^2-\alpha^2}
	\end{equation}
	where  $\varphi$  represents angle in $S^1$. The expression for winding number is as follows
	\begin{equation}
	2\pi m= \int_{0}^{2\pi\alpha} \psi^\prime ~d\xi= \frac{A_2-\beta\omega_2}{\beta^2-\alpha^2}\int d\xi
	\label{7}	\end{equation}
	We get a condition by setting $m_0=0$, i.e, absence of winding in the $t$ direction, given by
	\begin{equation}
	2\pi m_0 = \int_{0}^{2\pi\alpha}  \phi_0^\prime ~d\xi = \int_{0}^{2\pi\alpha} d\xi~ \left(\frac{A_0-q\alpha\omega_1}{1+r_1^2}+(q\alpha\omega_1+\beta\omega_0)\right)=0
	\label{8}\end{equation}
	The Lagrangian with parameters $r_0,r_1$ can be expressed as
	\begin{eqnarray}
	L&=& -\frac{1}{2}\Big[r_0^{\prime 2}(\beta^2-\alpha^2) + \frac{1}{r_0^2(\beta^2-\alpha^2)}\left(A_0-q\alpha\omega_1r_1^2\right)^2-\frac{\alpha^2r_0^2\omega_0^2}{\beta^2-\alpha^2}-r_1^{\prime 2}(\beta^2-\alpha^2)\nonumber\\&&- \frac{r_1^2}{\beta^2-\alpha^2}\left(\frac{A_1}{r_1^2}-q\alpha\omega_0\right)^2+\frac{r_1^2\alpha^2\omega_1^2}{\beta^2-\alpha^2}-\frac{A_2^2}{\beta^2-\alpha^2}+\frac{\alpha^2\omega_2^2}{\beta^2-\alpha^2}\nonumber\\&& -\frac{2q\alpha r_1^2}{\beta^2-\alpha^2}\left(\frac{\omega_0A_1}{r_1^2}+\frac{\omega_1A_0}{r_0^2}+q\alpha(\omega_1^2-\omega_0^2)\right)+ \Lambda(r_0^2-r_1^2-1)\Big]
	\end{eqnarray}
	The corresponding Hamiltonian then becomes
	\begin{eqnarray}
	H&=& \frac{1}{2}\Big[-r_0^{\prime 2}(\beta^2-\alpha^2) + \frac{1}{r_0^2(\beta^2-\alpha^2)}\left(A_0-q\alpha\omega_1r_1^2\right)^2-\frac{\alpha^2r_0^2\omega_0^2}{\beta^2-\alpha^2}+r_1^{\prime 2}(\beta^2-\alpha^2)\nonumber\\&&- \frac{r_1^2}{\beta^2-\alpha^2}\left(\frac{A_1}{r_1^2}-q\alpha\omega_0\right)^2+\frac{r_1^2\alpha^2\omega_1^2}{\beta^2-\alpha^2}-\frac{A_2^2}{\beta^2-\alpha^2}+\frac{\alpha^2\omega_2^2}{\beta^2-\alpha^2}\nonumber\\&& -\frac{2q\alpha r_1^2}{\beta^2-\alpha^2}\left(\frac{\omega_0A_1}{r_1^2}+\frac{\omega_1A_0}{r_0^2}+q\alpha(\omega_1^2-\omega_0^2)\right)\Big]
	\end{eqnarray}
	Calculating constraints, we find the first and second constraints takes the following form
	\begin{equation}
	(\beta^2-\alpha^2)(r_0^{\prime 2}-r_1^{\prime^2})+(\beta^2-\alpha^2)(r_0^2\phi_0^{\prime 2}-\phi_1^{\prime 2} r_1^2-\psi^{\prime 2})-\omega_0^2r_0^2+\omega_1^2r_1^2+\omega_2^2=0
	\label{3}\end{equation}
	\begin{equation}
	\omega_0A_0 +\omega_1A_1+\omega_2A_2=0
	\label{4}	\end{equation}  
	We consider a particular case here, i.e., $A_2=0$. Now, with the choice of $\omega_0$ and $\omega_1$ to be positive, the constants $A_0$ and $A_1$ can have opposite signs as depicted from the above equation.\\
	The conserved charges are:
	\begin{equation}
	E=T\int -\frac{d\xi}{\alpha (\beta^2-\alpha^2)}\left(A_0\beta+\alpha^2\omega_0+(1-q^2)\alpha^2r_1^2\omega_0+q\alpha A_1\right)
	\end{equation}	 
	\begin{equation}
	S=T\int \frac{d\xi}{\alpha (\beta^2-\alpha^2)}\left(A_1\beta-\frac{\alpha^2\omega_1 r_1^2}{1+r_1^2}-\frac{(1-q^2)\alpha^2r_1^4\omega_1}{1+r_1^2}+\frac{q\alpha A_0 r_1^2}{1+r_1^2}\right)
	\end{equation}
	\begin{equation}
	J=T\int\frac{d\xi}{\alpha (\beta^2-\alpha^2)}\left(A_2\beta-\alpha^2\omega_2\right)
	\end{equation}
	Imposing the condition, $r_0^2-r_1^2=1$ in (\ref{3}), the equation for $r_1$ can be written as
	\begin{eqnarray}
	(\beta^2-\alpha^2)^2r_1^{\prime 2} = (1+r_1^2)\Bigg(&&\frac{(A_0+q\alpha\omega_1r_1^2)^2}{1+r_1^2}-\frac{(A_1-q\alpha \omega_0r_1^2)^2}{r_1^2}-A_2^2\nonumber\\&&+ \alpha^2\omega_0^2(1+r_1^2)-\alpha^2\omega_1^2 r_1^2- \omega_2^2\alpha^2\Bigg)
	\label{5}\end{eqnarray}
	Solving the above equation results in finding the arc of the string for two finite values of $r_1$, which corresponds to the turning points of the string profile. Here we consider only those string solutions which doesn't reach the boundary, which demands the following condition $\omega_0^2<\omega_1^2$ to be satisfied.\\
	Now we express equation (\ref{5}) in terms of variable $v$ which is related to the global coordinate $\rho$ as follows
	\begin{equation}
	v=\frac{1}{1+2r_1^2}= \frac{1}{\cosh2\rho}, ~~~~~0\leq v\leq 1,
	\end{equation}  
	where we have used $r_1= \sinh\rho.$
	\begin{eqnarray}
	(\beta^2-\alpha^2)&&v^{\prime 2}= 2v \Bigg(4v^2 (1-v)\Big(A_0+q\alpha\omega_1\big(\frac{1-v}{2v}\big)\Big)^2 - 4v^2(1+v)\Big(A_1-q\alpha\omega_0\big(\frac{1-v}{2v}\big)\Big)^2\nonumber\\&&+\omega_0^2\alpha^2(1+v)^2(1-v)-\omega_1^2\alpha^2(1+v)(1-v)^2-2( A_2^2+\omega_2^2\alpha^2) v (1-v^2)\Bigg)
	\label{6}\end{eqnarray}
	We fix $\alpha=1$ in the above equation without affecting its prevalence, resulting in the equation (\ref{6}) taking the form
	\begin{equation}
	v^\prime = \frac{\sqrt{2vP(v)}}{1-\beta^2}
	\end{equation}	
	where 
	\begin{eqnarray}
	P(v)&=& v^3\left[-4A_0^2-4A_1^2+2(A_2^2+\omega_2^2)-(1+q^2)(\omega_0^2+\omega_1^2)+4q(A_0\omega_1-A_1\omega_0)\right]\nonumber\\&&+v^2\left[4A_0^2-4A_1^2-\omega_0^2+\omega_1^2-8A_0q\omega_1+q^2(\omega_0^2+3\omega_1^2)\right]\nonumber\\&&+~v[\omega_0^2+\omega_1^2-2(A_2^2+\omega_2^2)+4q(A_1\omega_0+A_0\omega_1)+q^2(\omega_0^2-3\omega_1^2)]\nonumber\\&&+(1-q^2)(\omega_0^2-\omega_1^2)
	\end{eqnarray}
	\begin{eqnarray}
	\equiv \Bigg[-4A_0^2-4A_1^2+2(A_2^2+\omega_2^2)&-&(1+q^2)(\omega_0^2+\omega_1^2)+4q(A_0\omega_1-A_1\omega_0)\Bigg] \nonumber\\&&\times\left((v-v_1)(v-v_2)(v-v_3)\right)
	\end{eqnarray}
	where $v_1,v_2,v_3$ are the three real roots of $P(v)=0$ to account for a consistent solution. Different conditions i.e, (\ref{7}),(\ref{8}),(\ref{4}) can be used to eliminate various constants by expressing one in terms of other therby reducing the number.\\
	Let us choose the constants $A_0, A_1$ in such a way that we get two positive real roots to $P(v)$ $0\leq v_2\leq v_3\leq 1$ and one negative real root $v_1\leq 0$. Now the product of the roots can be determined as
	\begin{eqnarray}
	a\equiv-4A_0^2-4A_1^2+2(A_2^2+\omega_2^2)&-&(1+q^2)(\omega_0^2+\omega_1^2)+4q(A_0\omega_1-A_1\omega_0)\nonumber\\&& = \frac{(1-q^2)(\omega_1^2-\omega_0^2)}{v_1 v_2 v_3}
	\end{eqnarray}
	We are interested in the range of the roots in which the polynomial $P(v)$ remains positive. From the property of a third-degree polynomial, the positive nature of the polynomial can be checked by looking at the sign of the parameter, here, $a$ being negative. We observe here that, for two positive roots $v_2\leq v \leq v_3$ and one negative root $v_1$, the parameter is found to be negative. i.e., $a<0$ thereby specifying the positive nature of the polynomial $P(v)$ between the two positive roots. So by fixing the values of the two constants $v_2,v_3$, we can find the other values $v_1, A_0, A_1$ in terms of $v_2,v_3$. The polynomial now can be written as 
	\begin{eqnarray}
	P(v)=\frac{(1-q^2)(\omega_1^2-\omega_0^2)}{v_1 v_2 v_3}(v-v_1)(v-v_2)(v-v_3) = -8A_1^2\frac{(v-v_1)(v-v_2)(v-v_3)}{(1-v_1)(1-v_2)(1-v_3)}\nonumber\\
	\end{eqnarray}
	where $v_1$ can be written in terms of $v_2,v_3$ as:
	\begin{equation}
	v_1= -\frac{v_2v_3}{v_2+v_3+v_2v_3\Big(\cfrac{\omega_0^2+\omega_1^2-2(A_2^2+\omega_2^2)+4q(A_1\omega_0+A_0\omega_1)+q^2(\omega_0^2-3\omega_1^2)}{(1-q^2)(\omega_0^2-\omega_1^2)}\Big)}
	\end{equation}
	Similarly, $A_0, A_1$ takes the following form 
	\begin{equation}
	(A_0-q\omega_1)^2 = \frac{(1-q^2)(\omega_0^2-\omega_1^2)}{8}\frac{(1+v_1)(1+v_2)(1+v_3)}{v_1v_2v_3}
	\end{equation}
	\begin{equation} A_1^2=\frac{(1-q^2)(\omega_0^2-\omega_1^2)}{8}\frac{(1-v_1)(1-v_2)(1-v_3)}{v_1v_2v_3}
	\end{equation}
	Here we can check the consistency of the choice of roots of the polynomial $P(v)$ from the non negative values of $A_0^2$ and $A_1^2$ i.e, $A_0^2\ge 0$ and $A_1^2 \ge 0$ for a solution satisfying $-1\leq v_1 \leq 0 \leq v_2 \leq v_3 \leq 1$.\\
	We glue together 2$N$ pieces of integrals between valley $(v_2)$ and spike $(v_3)$ in order to get string solutions for $N$-spikes. So, we need to replace all the integrals $\int d\xi$ by
	\begin{equation}
	\int d\xi = 2N (1-\beta^2)\int_{v_2}^{v_3} \frac{dv}{v^\prime}= \frac{2n}{\sqrt{-2a}}I_1
	\end{equation}  
	where $I_1$ is given in the Appendix in terms of elliptic integrals. With this substitution, the expression for winding number $m$ in (\ref{7}) takes the following form
	\begin{equation}
	m=\frac{\beta \omega_2-A_2}{\pi\sqrt{-2a}} NI_1
	\label{10}\end{equation}
	Using (\ref{10}) for $A_2$ in (\ref{4}), we find the following equation for $\omega_2$
	\begin{equation}
	\omega_0A_0+\omega_1A_1+\Big(\beta\omega_2-\frac{\pi m \sqrt{-2a}}{NI_1}\Big)=0
	\end{equation}
	We can eliminate one of the constants, say $\beta$ using (\ref{8}) as follows
	\begin{equation}
	2A_0I_5+q \alpha\omega_1I_1-2q\alpha\omega_1I_5+\omega_0\beta I_1=0
	\label{9} \end{equation}
	Now we will calculate the expression for conserved charges for solution with $N$ spikes as
	\begin{equation}
	\frac{\pi \mathcal{E}}{N}= \frac{1}{\sqrt{-2a}}\left(\beta A_0+\omega_0+q A_1\right)I_1+\frac{(1-q^2)\omega_0}{2\sqrt{-2a}}I_2,~~	\frac{\pi \mathcal{J}}{N}= \frac{\omega_2-A_2\beta}{\sqrt{-2a}}I_1
	\end{equation}
	\begin{eqnarray}
	\frac{\pi \mathcal{S}}{N}= -\frac{1}{\sqrt{-2a}}\Bigg(\big(A_1\beta+ qA_0-2q^2\omega_1\big)I_1-\frac{\omega_1}{2}I_2+\frac{q^2\omega_1}{2}I_3+2(\omega_1q^2-qA_0)I_5\Bigg)\nonumber\\ 
	\end{eqnarray}
	
	Where the integrals used are determined in terms of elliptic integrals in the Appendix. As already discussed in the earlier section, we use the following notation $E=2\pi T \mathcal{E},~~S=2\pi T \mathcal{S},~~J=2\pi T \mathcal{J}$ to compute the conserved charges.\\
	Now we want to find out the expression for the number of spikes. This can be calculated from $\theta$ which in turn can be obtained from the embedding coordinate $Y_1$ as
	$$Y_1= \sinh\rho e^{i\theta}, ~~~~~~~~~~\theta=\omega_1\tau +\int d\xi \phi_1^\prime$$
	From here we can find the number of spikes via $\Delta{\theta}=\cfrac{2\pi}{2n}$ at fixed $$t=\omega_0 \tau + \phi_0(\xi)$$
	Here $\Delta\theta$ is the angular separation i.e, the angle between a minimum and a maximum 
	\begin{eqnarray}
	\Delta\theta&=&\int d\theta = \int\frac{ d\xi}{\beta^2-\alpha^2} \left(\frac{A_1}{r_1^2}+\frac{A_0\omega_1}{\omega_0(1+r_1^2)}-q\omega_0+\frac{q\omega_1^2r_1^2}{\omega_0(1+r_1^2)}\right)\nonumber\\
	&&= -\frac{2}{\sqrt{-2a}}\left(A_1I_6+\frac{\omega_1}{\omega_0}A_0I_5-q\frac{\omega_1^2}{\omega_0}I_5+\frac{q}{2\omega_0}(\omega_1^2-\omega_0^2)I_1\right)
	\end{eqnarray}
	Due to the presence of angular momentum $J$ and winding $m$ in $S^1$, we would like to examine whether the spikes end in cusps in the presence of flux. In order to verify this, we ought to evaluate the derivative of $\rho$ with respect to $\theta$ at the minimun value of $v=v_2$ with fixed $t$ 
	\begin{equation}
	\frac{d\rho}{d\theta}\Bigg|_{v=v_2}= \frac{\rho^\prime d\xi}{\omega_1 d\tau+ \phi_1^\prime d\xi}\Bigg|_{v=v_2}
	\end{equation}
	Solving $dt=\omega_0d\tau+\phi_0^\prime d\xi=0$ for fixed $t$, and using the result in the above equation, we get
	\begin{equation}
	\frac{d\rho}{d\theta}\Bigg|_{v=v_2}= \frac{\rho^\prime}{\phi_1^\prime-\frac{\omega_1}{\omega_0} \phi_0^\prime}\Bigg|_{v=v_2}
	\end{equation}
	Evaluating the above expression we get
	\begin{equation}
	\frac{d\rho}{d\theta}\Bigg|_{v=v_2}= \frac{\sqrt{P(v)}}{\sqrt{2v}\sqrt{1-v^2}}~~\frac{\omega_0 \omega_1}{\omega_0^2+\omega_1^2}~~\cfrac{1}{\left(\frac{\omega_1^2-\omega_0^2}{\omega_0^2+\omega_1^2}-v\right)\left(q\omega_1+\frac{2vA_0}{1-v^2}\right)}\Bigg|_{v=v_2}
	\label{11}	
	\end{equation}
	\begin{figure}[t]
		
\includegraphics[scale=0.7]{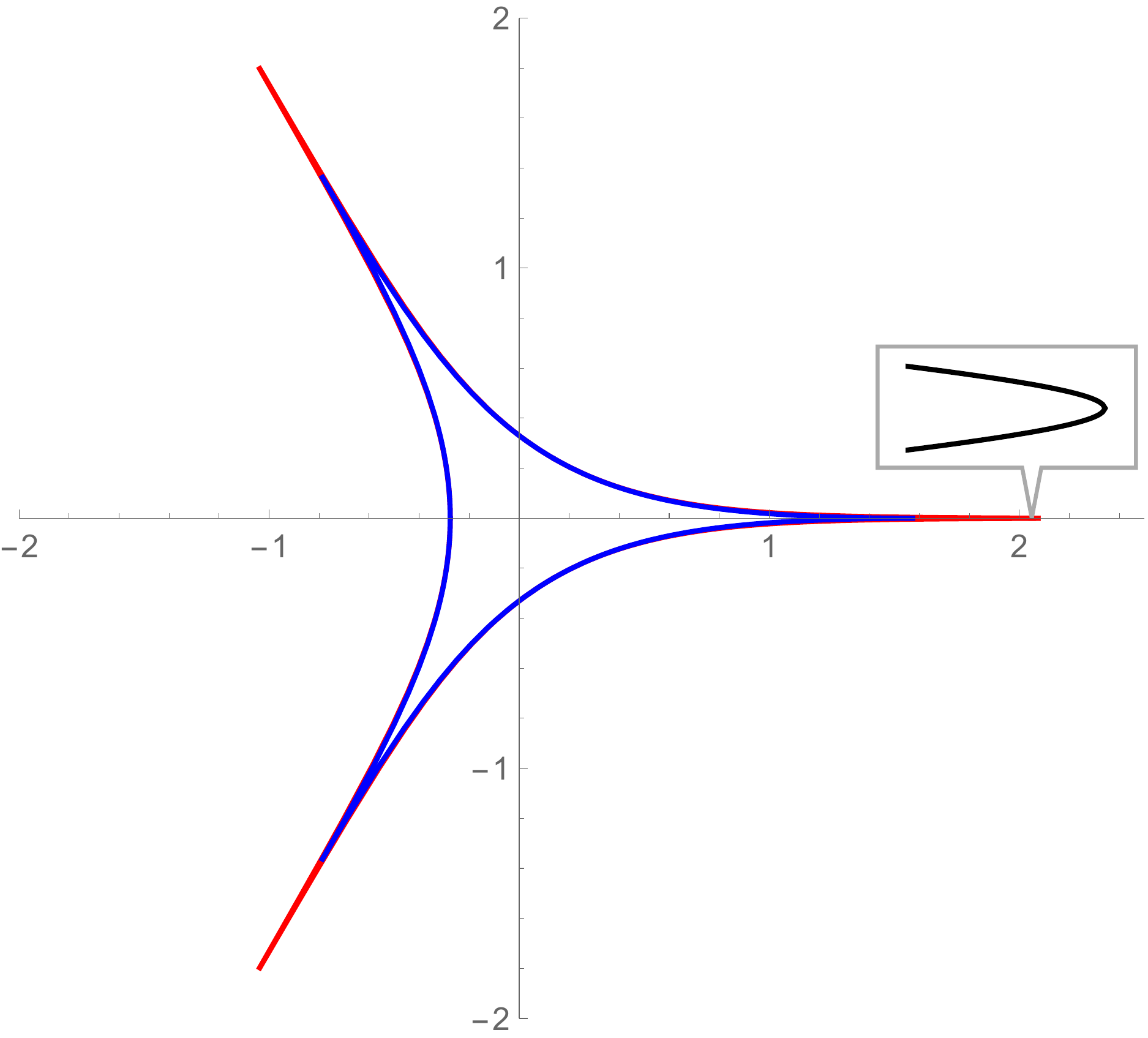}
		\caption{String profile in the presence of NS-NS flux for $N=3$ in plane polar coordinates ($\rho,\theta$) has been plotted for two different values of $q$, i.e, $q=0$ (red) and $q=0.1$ (blue). The two plots doesnot seem to vary much in this scale due to negligible difference in the q value but the rounding off nature is seen in both the cases. Here we have used $\rho= \sinh^{-1} \sqrt{\frac{1 - v}{2 v}}$ and eqn (\ref{thetamixed}) for $\theta$. The values of the parameters taken for $q=0.1$ are $v_1=-0.9527, v_2=0.0858, v_3=0.865$ and for $q=0$ are $v_1=-0.8698, v_2=0.04, v_3=0.0314$. We can clearly see the rounding-off nature of the spike here, i.e., the spikes do not end in cusp.}
	\end{figure}  
	
	We notice here that the denominator does not vanish in general for $P(v_2)=0$, which would have resulted in spikes. In particular it does vanish when $v_2=\frac{\omega_1^2-\omega_0^2}{\omega_0^2+\omega_1^2}$ but this case corresponds to the situation when motion is only in $AdS_3$. Thus, we claim that when the string moves or extends in $S^1$, the spikes at $v=v_2$ get rounded-off, i.e., they do not end in cusps. \\
	To illustrate the rounding of the spikes, we take $A_2=0$, keeping $m$ and $\omega_2$ non-zero. The expression for the angle $\theta$ in terms of $v_1,v_2,v_3$ at fixed t is given by
	\begin{eqnarray}
	\theta(v)=-\frac{\sqrt{(1-v_1)(1-v_2)(1-v_3)}}{2}\left(I_6+\frac{A_0^2}{A_1^2}I_5-\frac{q}{2}\frac{\omega_0}{A_1}\Big(\frac{A_0^2+A_1^2}{A_1^2}\Big)I_5\right)
	\label{thetamixed}\end{eqnarray}
	For completeness, we plot the string profile in figure 1. Here we impose the closedness condition by satisfying, $\Delta \theta=\cfrac{2\pi}{2N}$ where $N\in \mathbb{Z}$. The profile seems exactly similar to the original spiky string discussed in \cite{spiky1}. However, a more detailed analysis reveals that the spikes are rounded off instead of ending in cusps. The plot is parametrized by the flux parameter $q=0.1$. So it can be seen here that the property of rounding of the spikes with non-zero angular momentum `$J$' in $S^1$ remains intact in the presence of flux.
	We find the expression for winding $m$ by solving for $\beta$ and using this in (\ref{10}),
	\begin{eqnarray}
	m=&& \frac{NI_1 \omega_2}{\pi } \frac{\sqrt{v_1 v_2 v_3}}{\sqrt{2(1-q^2)(\omega_0^2-\omega_1^2)}}\Bigg(\frac {q\omega_1}{\omega_0}(2 I_5-1)\nonumber\\ &-&2\frac{I_5}{\omega_0 I_1}\left(\frac{\sqrt{(1-q^2)(\omega_0^2-\omega_1^2)}}{2\sqrt{2}\sqrt{ v_1 v_2 v_3} \sqrt{(1+v_1)(1+v_2)(1+v_3)}+q\omega_1}\right)\Bigg)
	\end{eqnarray}
	The above formalism is a particular case, i.e., it is based on the assumption ($A_2=0$); however, we can adopt the similar formalism for the general case, where $A_2\neq0$. It can be argued in the general case that the expression for the charges; Energy, Spin, Angular momentum, and winding number can be expressed in terms of independent parameters.
	
	\subsection{Special Cases}
	In the following section, we consider a couple of particular cases to the general solutions studied in the previous section. In the first case, we shall retrieve the original spiky string solution with mixed flux considered in section 3 in the limit when the string is constrained to move along $S^5$ direction. In the second case, we shall generalize ``$pp-$ wave limit" discussed in section 2 for the $N$-spike string in $AdS_3$. This can be achieved by considering the motion of spikes close to the boundary of $AdS_5$. 
		
	\subsubsection{No angular momentum in $S^1$}
	In this case we set $m=\nu=0$ which demands $A_2=0$ by the condition (\ref{7}). This when put in the constraint equation (\ref{4}), gives  $\omega_0A_0+\omega_1A_1=0$. This can also be written as 
	\begin{equation}
	A_0=\omega_1h,~~~~~~~~A_1=-\omega_0h
	\end{equation}
	The polynomial $P(v)$ takes the form
	\begin{eqnarray}
	P(v)&=& -v^3(1+(2h-q)^2                                                                                                                                    ) (\omega_0^2+\omega_1^2)\nonumber\\&& +v^2\left[(1+(2h-q)^2)(\omega_1^2-\omega_0^2)+(2q^2-4qh)(\omega_1^2+\omega_0^2)\right] \nonumber\\ &&+v\left[(1-q^2)(\omega_1^2+\omega_0^2)+ (4qh-2q^2)(\omega_1^2-\omega_0^2)\right]\nonumber\\&&+(1-q^2)(\omega_0^2-\omega_1^2)
	\end{eqnarray}
	with the appropriate choice for the roots being $(v_2\leq v_3)$
	\begin{equation}
	v_1=-\frac{q^2-2qh-\sqrt{1+(2h-q)^2}}{(1+(2h-q)^2},~ v_2=\frac{\omega_1^2-\omega_0^2}{\omega_1^2+\omega_0^2},~ v_3=\frac{q^2-2qh+\sqrt{1+(2h-q)^2}}{(1+(2h-q)^2}
	\label{12}\end{equation}
	If we interchange the values of $v_2$ with $v_3$, we obtain a kind of string profile for which the spikes appear at the minimum value of $\rho$. In this section, however, we deal with such kind of string profile, in which spikes correspond to the maximum value of $\rho$.  \\
	Here, it can be seen clearly that the expression, $\frac{d\rho}{d\theta}|_{v=v_2}$ blows up for $v=v_2$ indicating the presence of spikes. 
	As done in earlier section, we use the condition (\ref{9}) in order to eliminate the parameter $\beta$ 
	\begin{equation}
	\beta=\frac{2\omega_1(q-h)}{\omega_0}\frac{I_5}{I_1}-q\frac{\omega_1}{\omega_0}
	\end{equation}
	The expression for the conserved charges are as follows
	\begin{eqnarray}
	&&\mathcal{E}= \frac{N}{2\pi \omega_0 \sqrt{-2a}}\Bigg[4\omega_1^2h(q-h) I_5+\omega_0^2I_3-q^2\omega_0^2I_2-2qh(\omega_1^2+\omega_0^2)I_1\Bigg]\\
	&&\mathcal{S}=-\frac{N}{2\pi\sqrt{-2a}}\Bigg[4 \omega_1h^2I_5-\omega_1I_2-4q\omega_1h(2I_5-I_1)-2q^2\omega_1(2I_1-\frac{I_3}{2})\Bigg]\\&&\frac{\pi}{N}=\frac{2}{\omega_0\sqrt{-2a}}\Bigg[\omega_1^2hI_5-\omega_0^2hI_6-q\omega_1^2I_5+\frac{q}{2}(\omega_1^2-\omega_0^2)I_1\Bigg]
	\end{eqnarray}
	where
	\begin{equation}
	\sqrt{-2a}=\sqrt{\frac{2(\omega_1^2+\omega_0^2)(q^2-1)(1+(q-2h)^2)^2}{1+(1-q^2)(2h-q)^2}}
	\end{equation}
	The integrals, $I_n$ used above are expressed in the standard elliptic integrals form in the Appendix. Here, we are left with two independent parameters $\cfrac{\omega_0}{\omega_1},h$, which in turn, is needed to express energy as  $\mathcal{E}=\mathcal{E}(\mathcal{S},N).$
	We consider large  $\mathcal{S}$ limit ($\omega_1\to\omega_0$) which corresponds to small $v_2$ and find that the integrals $I_5$ and $I_6$ are finite in this limit. The number of spikes can now be calculated as
	\begin{equation}
	\frac{\pi}{N}= \sqrt{\frac{1+(1-q^2)(2h-q)^2}{(1+(q-2h)^2)(q^2-1)}}\Bigg(h(I_5-I_6)-q\omega_0^2I_5\Bigg)
	\end{equation}
	The charges scale as 
	\begin{eqnarray}
	\mathcal{S}&\simeq&\frac{N}{2\pi}\frac{\sqrt{q^2-1}\sqrt{1+(q-2h)^2(1-q^2)}}{q(2h-q)}\frac{1}{v_2}\nonumber\\
	&&\ln\frac{\mathcal{S}}{N}\simeq -\ln{v_2}
	\end{eqnarray}
	 and hence, the dispersion relation can be expressed as
	\begin{equation}
	\mathcal{E-S}\simeq~~-C(q)\frac{N}{2\pi} \ln v_2 ~~\simeq~~ C(q)\frac{N}{2\pi}\ln\frac{\mathcal{S}}{N}
	\end{equation}
	We can see from the above expression that the dispersion relation retains its log $S$ behavior at the leading order, similar to the one we encounter in the spiky string solution in the AdS background.

	\subsubsection{$AdS-pp-$wave limit}
	In order to obtain the solution in the ``$pp-$wave limit" which can also be achieved by taking $N\to\infty$, we consider the following scaling $(k=1,2,3)$
	\begin{equation}
	v_k\to \epsilon^2 v_k,~~~~~~\epsilon\to 0
	\label{19}\end{equation}
	It becomes apparent here that $\omega_2$ and $A_2$ ought to remain fixed in this scaling limit, while the constants $A_0, A_1$ scale in the following manner 
	\begin{equation}
	A_0=a_0\left(\frac{1}{\epsilon^2}+\frac{v_1+v_2+v_3}{2}+...\right)+q\omega_1,~~~A_1=-a_0\left(\frac{1}{\epsilon^2}-\frac{v_1+v_2+v_3}{2}+...\right)
	\end{equation}
	where $$a_0=\sqrt\frac{(1-q^2)(\omega_0^2-\omega_1^2)}{8v_1v_2v_3\epsilon^2}$$
	Also, we want  $\omega_0, \omega_1$ to scale as
	\begin{equation}
	\frac{\omega_0-\omega_1}{\omega_0}=r\epsilon^2
	\end{equation}
	where r is finite. We use $\omega_0^2-\omega_1^2=8\epsilon^2a_0^2v_1v_2v_3$ from which the expression for $r$ is found to be $r=\frac{4a_0^2}{\omega_0^2}v_1v_2v_3$.
	Other expressions can be calculated as:
	\begin{eqnarray}
	&&2(A_2+\omega_2^2)~-~(1+q^2)(\omega_0^2+\omega_1^2)\nonumber\\&&=\frac{(1-q^2)(\omega_1^2-\omega_0^2)}{v_1 v_2 v_3}+4q(A_0\omega_1+A_1\omega_0)+4(A_0-q\omega_1)^2+4A_1^2-4q^2\omega^2\nonumber\\&&
	=8a_0^2(v_1v_2+v_2v_3+v_3v_1)+2qa_0(v_1+v_2+v_3)(\omega_0+\omega_1)-4q\omega_0a_0r\\&&~~~~~~~~~~~~~~~~~~~~~~~~~~~~~~~~a=-\frac{8a_0^2}{\epsilon^4}
	\end{eqnarray}
	The constraint equation takes the form
	\begin{equation}
	a_0r+a_0(v_1+v_2+v_3)+\frac{\omega_2}{\omega_0}A_2+q\omega_1=0
	\end{equation}
	The above equation can be used to eliminate one constant. We find that under (\ref{19}), different integrals scale as 
	\begin{equation}
	I_1\to\frac{1}{\epsilon^2}I_1,~~~I_2\to\frac{1}{\epsilon^4}I_+,~~~I_3\to\frac{1}{\epsilon^4}I_+,~~~I_5\to I_4,~~~I_6\to I_4
	\end{equation}
	where $I_+=\frac{1}{2}(I_2+I_3)$. We get the expression for $\beta$ from the following constraint equation
	\begin{equation}
	2a_0I_4+\omega_0\beta I_1+q\omega_1I_1=0
	\end{equation}	
	The form of equation for $v$ remains intact in this limit
	\begin{equation}
	v^\prime=\frac{\sqrt{2vP(v)}}{1-\beta^2},~~~~~P(v)=-8a_0^2(v-v_1)(v-v_2)(v-v_3)
	\end{equation}
	The polynomial $P(v)$ can be expressed as
	\begin{eqnarray}
	P(v)&=&v^3\big[-4A_0^2-4A_1^2+4q(A_0\omega_1-A_1\omega_0)\big]+v^2\big[4A_0^2-4A_1^2-8qA_0\omega_1\big]\nonumber\\&+&v\big[4q(A_1\omega_0+A_0\omega_1)+(1-q^2)(\omega_0^2+\omega_1^2)\big]+(1-q^2)(\omega_0^2-\omega_1^2)
	\end{eqnarray}
	We will introduce some new parameters
	\begin{equation}
	\omega_\pm=\frac{\omega_1\pm\omega_0}{2},~~~~~~~~~~~~~~~~A_\mp=\frac{A_1\pm A_0}{2}
	\end{equation}
	
	we observe that they scale as 
	\begin{equation}
	A_-\sim\mathcal{O}(\epsilon^0),~~~~~A_+\sim\frac{1}{\epsilon^2},~~~~~\omega_-\sim(\epsilon^2),~~~~~\omega_+\sim\mathcal{O}(\epsilon^0)
	\end{equation}
	Under above scaling limit, the polynomial $P(v)$ scale as
	\begin{eqnarray}
	&&P(v)=-2A_+^2v^3~-~4[A_+A_-+qA_+\omega_+]v^2\nonumber\\&&+2\left[(1-q^2)\omega_+^2-\omega^2-A_\alpha^2-2q(A_-\omega_+-A_+\omega_-) \right]v-4(1-q^2)\omega_+\omega_-
	\end{eqnarray}
	The expression for the number of spikes $N$ and winding number $m$ is as follows
	\begin{equation}
	N=\frac{\pi\omega_0^2}{2~a_0~ v_1 v_2v_3}\left(a_0I_4-q\omega_0\right)\frac{1}{\epsilon^2},~~~\frac{m}{N}=-\frac{1}{4\pi a_0\omega_0}\left[(2a_0I_4+q\omega_1I_1)\omega_2+A_2\omega_0 I_1\right].
	\end{equation}
	In the above expression, we notice that the number of spikes rise to infinity while the ratio  $\cfrac{m}{N}$ remains finite in this scaling limit.
	The scaling limit (\ref{19}) when performed on the conserved charges, they take the form
	\begin{equation}
	\frac{\mathcal{E}}{N}\sim\frac{1}{\epsilon^2},~~~	\frac{\mathcal{S}}{N}\sim\frac{1}{\epsilon^2},~~~	\frac{\mathcal{J}}{N}=finite
	\end{equation}
	we also find that 
	$\cfrac{\mathcal{E-S}}{N}\sim \cfrac{\mathcal{E+S}}{N^2}\sim\cfrac{\mathcal{J}}{N}\sim\cfrac{m}{N}=$ stays finite.
	It can be seen here that one of the property of the scaling limit introduced above is merely to reduce the number of independent parameters.

	\section{Spiky string in $AdS-pp-$wave$\times S^1$ with flux}
	Here we generalize the solution of the periodic spike to its motion in $S^5$. We take the same limit as taken earlier to get $AdS-pp-$wave $\times S^5$ (the limit has no impact on $S^5$) and the metric can be given by
	\begin{eqnarray}
	ds^2&=&\frac{1}{z^2}\left(2dx_+dx_--\mu^2z^2dx_+^2+dz^2\right)+d\alpha^2\nonumber\\
	&&B_{x_+x_-}=\frac{q}{z^2}dx_+\wedge dx_-
	\end{eqnarray}
	where $\alpha$ pamametrizes a maximal circle $S^1$ and represent an angle of period $2\phi$. We find the equation of motion and Virasoro constraints for the strings moving in this background in the same way we did in our previous section . We pursue the following ansatz by analogy with the discussion in the preceding section.
	\begin{equation}
	x_\pm=\omega_\pm \tau+\phi_\pm(\xi),~~~\alpha=\omega\tau+\phi_\alpha(\xi),~~~z=z(\xi),~~~\xi=\sigma+\beta\tau
	\end{equation}
	The solution we get using this ansatz corresponds to a rigid string not only moving along $x=\frac{1}{\sqrt{2}}(x_++x_-)$ but wraps $S^1$ and move along it. The equations of motion for $x_\pm$ and $\alpha$ can be integrated as 
	\begin{eqnarray}
	\phi_\alpha^\prime&=&\frac{1}{1-\beta^2}(A_\alpha+\beta\omega),~~~~~\phi_+^\prime=\frac{1}{1-\beta^2}((\beta+q)\omega_++A_+z^2),\nonumber\\&&\phi_-^\prime=\frac{1}{1-\beta^2}(\beta\omega_-+A_-z^2+\mu^2A_+z^4+q(z^2\mu^2\omega_+-\omega_-))
	\end{eqnarray}
	where $A_\pm$ and $A_\alpha$ are constants of integration. First and second conformal gauge constraints can be written as
	\begin{equation}
	(1-\beta^2)\left(-\mu^2\phi_+^2+\frac{2}{z^2}\phi_+^\prime\phi_-^\prime+\frac{z^{\prime 2}}{z^2}+\phi_\alpha^{\prime 2}\right)+\omega^2+\frac{2}{z^2}\omega_+\omega_--\mu^2\omega_+^2=0
	\label{13}\end{equation}
	\begin{equation}
	\beta\left(-\mu^2\phi_+^2+\frac{2}{z^2}\phi_+^\prime\phi_-^\prime+\frac{z^{\prime 2}}{z^2}+\phi_\alpha^{\prime 2}\right)=\mu^2\omega_+\phi_+^\prime-\frac{\omega_+}{z^2}\phi_-^\prime-\frac{\omega_-\phi_+^\prime}{z^2}-\omega\phi_\alpha^\prime 
	\label{18}\end{equation}
	Using (\ref{18}) in (\ref{13}) we get 
	\begin{equation}
	A_-\omega_++A_+\omega_-+\omega A_\alpha=0
	\end{equation}
	\begin{eqnarray}
	(1-\beta^2)z^{\prime 2}&=&-\mu^2A_+^2z^6-2(A_+A_-+q\mu^2\omega_+A_+)z^4-2(1-q^2)\omega_+\omega_-\nonumber \\ &&+\left((1-q^2)\mu^2\omega_+^2-\omega^2-A_\alpha^2-2q(A_-\omega_+-A_+\omega_-)\right)z^2
	\end{eqnarray}
	Rewriting the second constraint by substituting $z^2=v$ as
	\begin{equation}
	v^\prime=\frac{\sqrt{2vP(v)}}{1-\beta^2}
	\end{equation}
	\begin{eqnarray}
	P(v)&=&-2\mu^2A_+^2v^3-4(A_+A_-+q\mu^2\omega_+A_+)v^2-4(1-q^2)\omega_+\omega_-\nonumber \\ &&+2\left((1-q^2)\mu^2\omega_+^2-\omega^2-A_\alpha^2-2q(A_-\omega_+-A_+\omega_-)\right)v
	\end{eqnarray}
	We plot this to get the shape of the string $z(u)$ and hence $x_\pm(\sigma, \tau)$. As can be seen clearly, we get identical equation by taking the limit discussed in the previous section. This generalizes the similarity between spiky string solution in $AdS_3$ and periodic spikes in $AdS-pp-$ wave background in the presence of flux. We can obtain other interesting string solutions using this ansatz.
	
	\section{Conclusions and Outlook}
	This paper studied the rigid rotating spiky string solutions in $AdS_3\times S^1$ background supported by NS-NS flux. We concentrate in particular on different limits of the spiky string solutions in $AdS_3$ in the presence of flux and generalization to it. The first limit taken into account is the large $N$ limit of the N-spike string in $AdS_3$ background with NS-NS flux. We find conserved charges to obtain the relation between them and observe exact resemblance with the expression to charges for a rigid string with periodic spikes in $AdS_3-pp-$ wave background. So the solutions we get with large $N$ limit can also be described by a particular spiky string solution moving in an $AdS-pp-$ wave background. The second part of this paper deals with two separate sections. In first section we study the addition of angular momentum $J$ (and winding m) in a maximal circle $S^1\subset S^5$ to $AdS_3$ background in the presence of flux. Here we use the conformal gauge to obtain the string solution. Although the shape of the string profile in presence of flux remains similar to the original spiky string profile, a more detailed analysis, however, shows a peculiar behavior of the spikes, which was no longer identical to the original one. The spikes do not end in cusps; instead are found rounded-off . This peculiarity arises due to the presence of the non-zero angular momentum $J$ and winding $m$ in $S^1$. The second section presents couple of cases, one is to see whether we retrieve the expressions for charges for the original spiky string with mixed flux as discussed in \cite{Banerjee:2019puc} when we take the case of zero angular momentum. Another one corresponds to taking the $pp-$ wave limit, which can also be achieved by taking $N\to \infty$. With this scaling limit we find that the ratios $\cfrac{\mathcal{E+S}}{N^2},\cfrac{\mathcal{E-S}}{N},\cfrac
	{\mathcal{J}}{N},\cfrac{m}{N}$ stays finite. It can be seen here that one of the property of this scaling limit is merely to reduce the number of independent parameters.
	Lastly, we generalize the periodic spike solution discussed in section 2 to the case of spiky sting moving in $S^5$. The expression we get for the polynomial $P(v)$ is similar to the one we get by applying $pp-$wave limit to spiky string solution in $AdS_3$ with mixed flux in the presence of angular momentum in $S^1$ discussed in section (4.2). This limit corresponds to a particular thermodynamic limit of the SL(2) spin chain on the gauge theory side.\\
	Moving forward to future studies, there are many different issues that one can study in a more significant way. The first follow-up could be to study the behavior of the spiky string in $AdS_3 \times S^3$ in the presence of flux both along $AdS_3$ and $S^3$ direction. One can check how the presence of flux along $S^3$ direction changes the string profile. Another issue for study may include D-string, which couples to both NS-NS and RR fluxes.  We would get back with some of these issues in the future.
	
	\subsection*{Acknowledgements}
	It is a pleasure to thank Manoranjan Samal for his discussion and valuable comments on the paper. One of us (RRN) would like to thank the financial support from
	Department of Science and Technology (DST), Ministry of Science and Technology,
	Government of India under the Women Scientist Scheme A (WOS-A) (grant no.
	SR/WOS-A/PM-62/2017). 
	
	\subsection*{Appendix 1}
	
	In the large $N$ limit, the expression for the Energy-Spin dispersion relation of the spiky string solution is expressed as:
	
	\begin{eqnarray}
		\frac{\mathcal{E-S}}{N} &&\to \frac{1}{2\pi v_0} \int_{v_1}^{(1-q)v_0}\frac{dv}{v^2}\frac{\sqrt{v(v-v_1)}~(v+qv_0)^2}{\sqrt{v_0^2-v^2-q^2v_0^2-2qvv_0}}\nonumber \\
		&+&\frac{1}{2\pi v_0}\int_{v_1}^{(1-q)v_0} dv \left(1-\frac{v_1}{v}\right)\frac{\sqrt{v_0^2-v^2-q^2v_0^2-2qvv_0}}{\sqrt{v(v-v_1)}}\nonumber \\
		&-&\frac{q}{2\pi}\int_{v_1}^{(1-q)v_0}\frac{dv}{v^2}\frac{\sqrt{v(v-v_1)}~(v+qv_0)}{\sqrt{v_0^2-v^2-q^2v_0^2-2qvv_0}},
	\label{20}\end{eqnarray}
	\begin{eqnarray}
		\frac{\mathcal{E+S}}{N^2}&&\to \frac{1}{2\pi^2}\int_{v_1}^{(1-q)v_0}\frac{dv}{v}\frac{\sqrt{v_0^2-v^2-q^2v_0^2-2qvv_0}}{\sqrt{v(v-v_1)}}\times\nonumber\\
		&&\int_{v_1}^{(1-q)v_0}\frac{dv^\prime}{v_0}\left(1+q\frac{v_0}{v^\prime}\right)\frac{\sqrt{v^\prime(v^\prime-v_1)}}{\sqrt{v_0^2-v^{\prime 2}-q^2v_0^2-2qv^\prime v_0}}.\nonumber\\
	\label{21}\end{eqnarray}

	Expression for the conserved quantities and anuglar separation between the spikes of periodic spike solution in $AdS-pp-$wave background are as follows:
	
	\begin{equation}
		P_-=~\frac{2T}{\mu z_0^2}~\int_{z_1}^{z_0\sqrt{1-q}}dz \frac{1}{z^2}\sqrt{\frac{((1+q)z_0^2+z^2)((1-q)z_0^2-z^2)}{z^2-z_1^2}},
	\label{22}\end{equation}
	\begin{eqnarray}
		P_+&=&~-\frac{\mu T}{z_0^2}~\int_{z_1}^{z_0\sqrt{1-q}}dz\frac{(z^2+z_0^2q)^2}{z^2}\sqrt{\frac{z^2-z_1^2}{(z^2+(1+q)z_0^2)((1-q)z_0^2-z^2)}}\nonumber\\&-&\frac{\mu T}{z_0^2}~\int_{z_1}^{z_0\sqrt{1-q}}dz\left(1-\frac{z_1^2}{2z^2}\right)\sqrt{\frac{(z^2+(1+q)z_0^2)((1-q)z_0^2-z^2)}{z^2-z_1^2}}\nonumber\\&+&\mu qT \int_{z_1}^{z_0\sqrt{1-q}}dz\left(1+\frac{qz_0^2}{z^2}\right)\sqrt{\frac{z^2-z_1^2}{(z^2+(1+q)z_0^2)((1-q)z_0^2-z^2)}}.
	\label{23}\end{eqnarray}
	where $z_0$ and  $z_1$ represents position of valley and spike respectively.
	
	\begin{eqnarray}
		\Delta x_-&=&\int d\sigma=-\eta_0^2\int d\xi\nonumber\\&&=-2\mu \int_{z_0\sqrt{1-q}}^{z_1}dz (z^2+qz_0^2)\sqrt{\frac{z^2-z_1^2}{(z^2+(1+q)z_0^2)((1-q)z_0^2-z^2)}}
	\end{eqnarray}
	Some of the integrals used in the calculations and their results:
	\begin{equation*}
	I_1= \int_{v_2}^{v_3} dv \frac{1}{\sqrt{-v(v-v_1)(v-v_2)(v-v_3)}},~~ I_2= \int_{v_2}^{v_3} dv \frac{1-v}{v\sqrt{-v(v-v_1)(v-v_2)(v-v_3)}}
	\end{equation*}
	\begin{equation}
	 I_3= \int_{v_2}^{v_3} dv \frac{1+v}{v\sqrt{-v(v-v_1)(v-v_2)(v-v_3)}}
	\end{equation}
	The above integrals can be computed as 
	\begin{eqnarray}
	&&I_1=\frac{2}{\sqrt{v_3(v_2-v_1)}} K[r],\\ &&I_2=-\frac{2}{v_1v_2}\sqrt{\frac{v_2-v_1}{v_3}}E[r]+\frac{2(1/v_1-1)}{v_3(v_2-v_1)}K[r]\\
	&&I_3=-\frac{2}{v_1v_2}\sqrt{\frac{v_2-v_1}{v_3}}E[r]+\frac{2(1/v_1+1)}{v_3(v_2-v_1)}K[r]
	\end{eqnarray}

	 \begin{eqnarray}
	 I_4&=&\int dv \frac{v}{\sqrt{-v(v-v_1)(v-v_2)(v-v_3)}}\nonumber\\
	 &&= \frac{-2}{\sqrt{v_3(v_2-v_1)}}\Bigg(v_1 F\left[\phi,r\right] +(v_3-v_1)\Pi\left[\frac{v_3-v_2}{v_1-v_2},\phi,r\right]\Bigg)
	 \end{eqnarray}
	 \begin{eqnarray}
	 I_5&=&\int dv \frac{v}{(1+v)\sqrt{-v(v-v_1)(v-v_2)(v-v_3)}}\nonumber\\
	 &&= \frac{-2}{(1+v_1)(1+v_3)}\frac{1}{\sqrt{v_3(v_2-v_1)}}\Bigg(v_1(1+v_3)F\left[\phi,r\right]\nonumber\\ &&+(v_3-v_1)\Pi\left[\frac{(1+v_1)(v_2-v_3)}{(v_2-v_1)(1+v_3)},\phi,r\right]\Bigg)
	 \end{eqnarray}
	 \begin{eqnarray}
	 I_6&=&\int dv \frac{v}{(1-v)\sqrt{-v(v-v_1)(v-v_2)(v-v_3)}}\nonumber\\
	 &&= \frac{-2}{(v_1-1)(v_3-1)}\frac{1}{\sqrt{v_3(v_2-v_1)}}\Bigg(v_1(v_3-1)F\left[\phi,r\right]\nonumber \\&&+(v_1-v_3)\Pi\left[\frac{(v_1-1)(v_2-v_3)}{(v_2-v_1)(v_3-1)},\phi,r\right]\Bigg)
	 \end{eqnarray}
	 where $$r=\sqrt{\frac{v_1(v_2-v_3)}{v_3(v_2-v_1)}},~~~~\phi=\sin^{-1}\sqrt{\frac{(v_2-v_1)(v-v_3)}{(v-v_1)(v_2-v_3)}}$$
	 \bibliographystyle{jhep}

\end{document}